\documentclass[11pt]{article}
\usepackage{moriond,epsfig,graphicx,subfigure}

\def\be{\begin{equation}}
\def\ee{\end{equation}}
\def\bea{\begin{eqnarray}}
\def\eea{\end{eqnarray}}

\begin{document}
\vspace*{4cm}
\title{THE BEAM-GAS METHOD FOR LUMINOSITY MEASUREMENT AT LHCb}

\author{ P. HOPCHEV }

\address{Laboratoire d'Annecy-le-vieux de Physique des Particules,\\ France}

\maketitle\abstracts{
The high resolution of the LHCb vertex detector makes it possible to 
perform precise measurements of the vertex positions of beam-gas and 
beam-beam interactions. With these measurements beam parameters such as 
width and position can be measured. A novel method for determining the 
absolute luminosity at the LHC using the directly measured beam 
parameters is presented. The data taken in 2009 is used to illustrate 
the procedure.}

\section{Plans for luminosity measurement in LHCb}
Luminosity is a fundamental accelerator characteristic, related to the amount of 
collisions produced. The rate of a certain process can be expressed as the product 
of the the cross-section for that process and the instantaneous luminosity.
LHCb is a forward spectrometer at the LHC optimised for precise studies of heavy 
flavour decays~\cite{lhcb}. In addition, the unique pseudo-rapidity 
coverage ($2 < y < 5$) of the detector will allow interesting measurements of 
production cross-sections in a hitherto unexplored kinematic region. For these 
measurements, good knowledge of the luminosity is essential.
In LHCb several methods for measuring the luminosity are under investigation.
\begin{enumerate}
\item Beam imaging with gas~\cite{massi1}, which is described in this paper
\item van der Meer scan, which consists of moving the two beams agains each other, 
while measuring the interaction rate
\item Several indirect methods using processes with well known cross-sections, e.g.
Z-boson and elastic diphoton dimuon production
\end{enumerate}

\section{Beam-gas luminosity method}

In a circular collider the luminosity for 2 colliding bunches of particles can be
expressed in the following way \cite{lum_formula}:

\begin{equation}
L ~ = ~ f N_{1}N_{2} ~ 2c ~ \cos^{2}(\phi/2) ~ \int{\rho_{1}(x,t)\rho_{2}(x,t) ~ d^{3}xdt}
\label{eq:overlapInt}
\end{equation}

\vspace{0.4cm}
\noindent where $f$ is the revolution frequency of the two counter-rotating bunches
travelling with the speed of light $c$, $N_{1,2}$ are the bunch intensities, 
$\phi$ is the beams crossing angle and $\rho_{1,2}$ are the time- and space-dependent
bunch densities. The integral in Eq.~\ref{eq:overlapInt} is known as the beam overlap
integral.

The beam-gas luminosity method is based on the detection of beam-gas vertices. 
The position of the beam-gas interactions can be used to measure the beam angles, 
profiles and relative positions. At a first approximation we can neglect possible
phase shifts and do not consider effects from the longitudinal shape of the bunches.
Then, having the transverse shapes of the bunches, we can calculate the overlap integral. 
The second important ingredient are the bunch intensities, knowledge of which comes 
from instrumentation installed and operated by the accelerator team.

Once the absolute luminosity is measured with satisfactory precision we will calibrate 
a reference cross-section and dedicated ``lumi counters'' in order to propagate the 
knowledge of the absolute scale. Currently the beam-gas luminosity method is being 
applied for a first time.

\section{VELO vertex reconstruction}
\begin{figure}[!h]
\begin{center}
\includegraphics[width=0.8\textwidth]{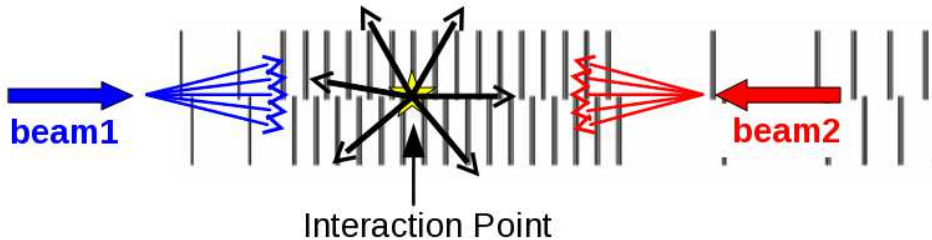}
\end{center}
\caption{
  A simplified sketch of the VELO, including the 2 pile-up stations on the left.
  The colour arrows indicate the direction of the two LHC beams and example 
  trajectories of the collision products from beam-gas interactions.
  Only the products of the beam1-gas interactions fly into the acceptance of 
  the LHCb spectrometer.
  \label{fig:velo_bg} }
\end{figure}

The LHCb vertex detector, the VELO, is located around the LHC interaction 
point 8 and can be used to reconstruct interaction vertices arising from 
beam-beam collisions and collisions between the beam protons and atoms of
the residual gas in the beam-pipe. The VELO consists of 21 stations of
radial and azimuthal silicon-strip sensors, see Fig.~\ref{fig:velo_bg}.
In addition two backward stations, the so called 'pile-up' system, are
used for providing a signal at the earliest level of the trigger system.

In November 2009 LHC delivered its first proton-proton collisions at center of mass 
energy equal to $900 GeV$. In the following weeks several million collision events were 
recorded by LHCb. The events were triggered by activity in the calorimeter system 
or from significant activity in the pile-up system in the backward region of the VELO.

\begin{figure}[!h]
\begin{center}
\includegraphics[width=\textwidth]{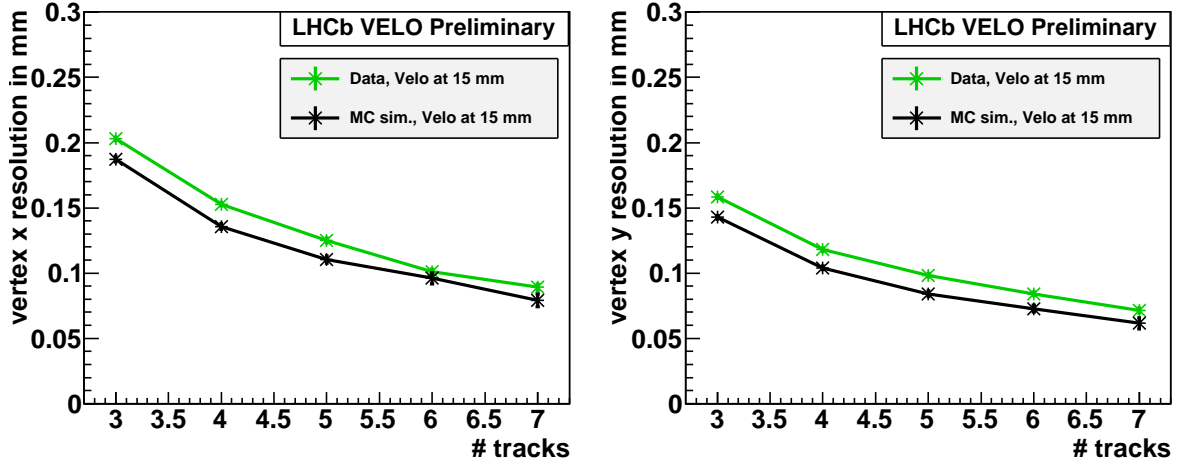}
\end{center}
\caption{
  Preliminary estimate of the VELO vertex resolution in the 2009 runs.
  The resolution in the transverse directions (x and y) for beam-beam interactions are shown
  as function of the number of tracks per reconstructed vertex.
  \label{fig:velo_res} }
\end{figure}

The vertex resolution for beam-beam interactions has been estimated with the data collected
in 2009. Preliminary results are shown in Fig.~\ref{fig:velo_res}. In 2009 the VELO was not
fully closed and each of the two halves was retracted by 15 mm in the horizontal direction
(along the x-axis). This leads to a worse resolution in x. For beam-gas interactions outside 
the luminous region we also take into account the dependence of the vertex resolution on the 
position of the vertex along the beam direction.

\section{Measured beam properties}
\begin{figure}[!ht]
\centering
\subfigure[]{
\includegraphics[width=0.45\textwidth]{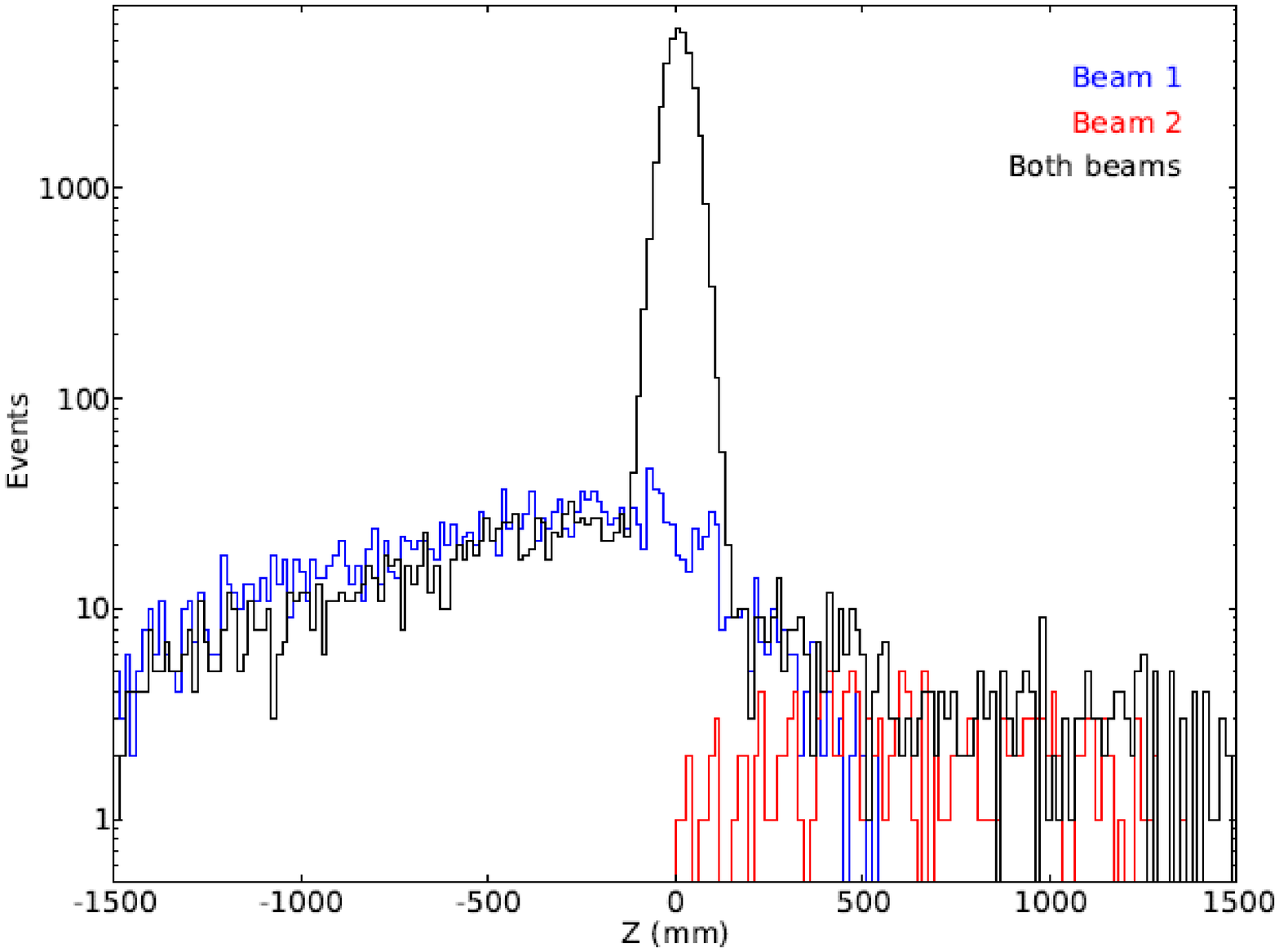}
\label{fig:vtc_z_pos}
}
\subfigure[]{
\includegraphics[width=0.5\textwidth]{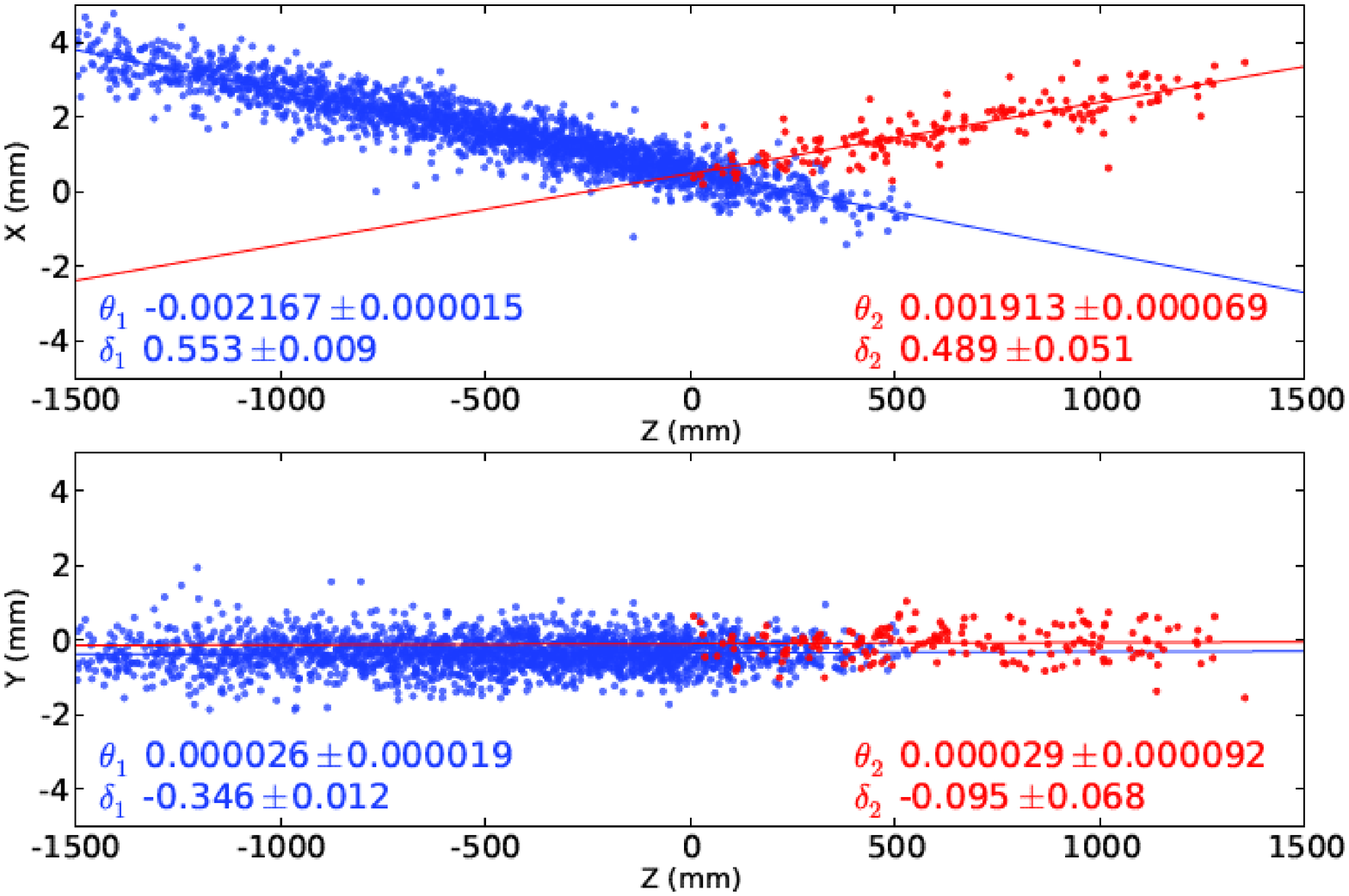}
\label{fig:bslopes}
}
\label{fig:vtc_positions}
\caption[]{Positions of reconstructed beam-gas vertices.
  \subref{fig:vtc_z_pos} Distributions of the z-position of the reconstructed beam-gas 
  (blue and red) and beam-beam (black) vertices. The asymmetry in the number of 
  reconstructed beam1-gas and beam2-gas interactions is due to the different trigger efficiency.
  \subref{fig:bslopes} Positions in the horizontal (x-z) and vertical (y-z) planes of the reconstructed 
  vertices from beam-gas interactions. The observed crossing angle in the x-z plane
  is due to the LHCb dipole magnet and is in agreement with the expected value.
}
\end{figure}

The VELO is positioned very close to the beam-axis which determines its very good
acceptance for beam-gas events along a wide range in z, where z is the coordinate 
measured parallel to the beam axis, see Fig.~\ref{fig:vtc_z_pos}.
The x-z and y-z projections of the positions of the reconstructed beam-gas vertices
can be used to determine the beams slopes and widths. Fig.~\ref{fig:bslopes} shows
the measured beam slopes during one of the 2009 runs.

\begin{figure}[!h]
\begin{center}
\includegraphics[width=\textwidth]{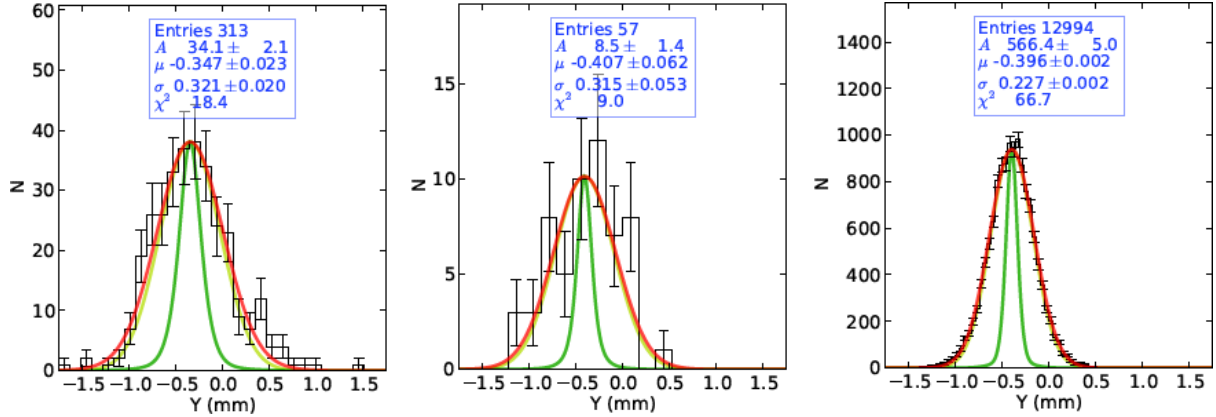}
\end{center}
\caption{
  Measured bunch and luminous-region profiles in the vertices (y-) direction. From left to
  right the plots correspond to bunch1, bunch2 and their luminous region. The green line 
  represents the vertex resolution, the red shows the observed size and the yellow - 
  the size after deconvolving the vertex resolution.
  \label{fig:bsizes} }
\end{figure}

\begin{figure}[!h]
\begin{center}
\includegraphics[width=\textwidth]{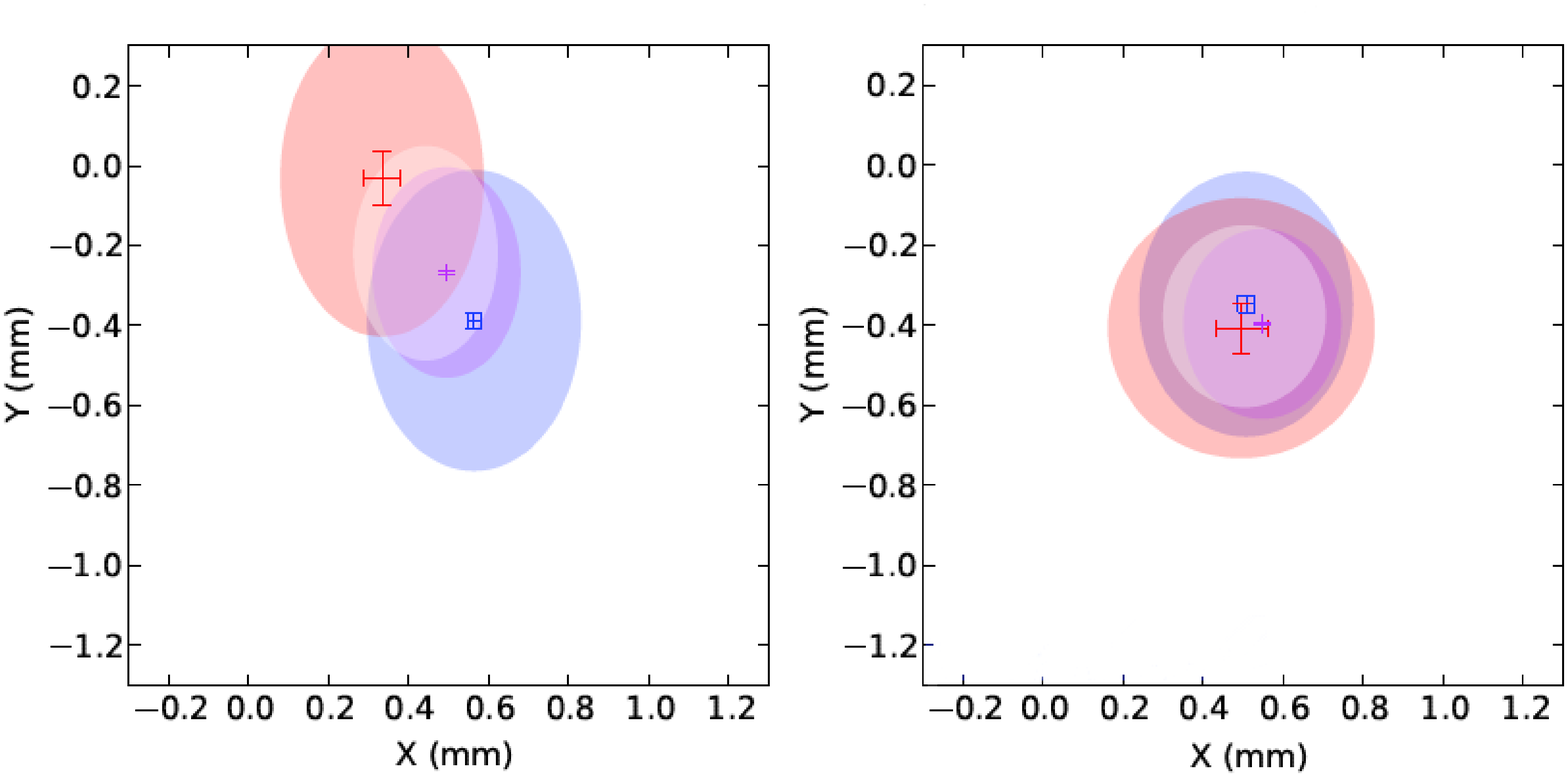}
\end{center}
\caption{
  Observed beam positions and sizes at z = 0 before and after beam adjustments (mini scan). 
  The colors have the following meaning: blue - beam1, red - beam2, light grey - predicted 
  luminous region, purple - measured luminous region.
  \label{fig:overlapp} }
\end{figure}

The measured beam and luminous-region sizes in the y-direction are shown in Fig.~\ref{fig:bsizes}.
One of the two colliding bunch pairs has been used to demonstrate the measurement.
The overlap of the the two beams in the interaction point is important for 
optimizing the luminosity. Using the beams directions and sizes one can also make a 
prediction about the collision region and learn more about the systematic
effects which have impact on its position and shape, Fig.~\ref{fig:overlapp}.

\section{Summary and prospects}

The presented preliminary studies show the feasibility of the beam-gas luminosity method.
For 2009 data the expected precision on the absolute luminosity is about $20\%$, 
decomposed roughly into $10\%$ from the measurement of the beam overlap and $15\%$ 
from the measurement of the beam intensities \footnote{The final analysis of the 2009 luminosity using
this method in fact achieved a relative precision of 15\% \cite{vladik}}.
It has been shown that the vertex resolution plays a small, but not negligible role in the 
determination the sizes of the beams. In 2010 more extensive studies will be possible 
allowing better estimate of the systematics and the higher amount of data will result in a 
decreased uncertainty. Considerable effort has been put into providing more 
precise beam intensity measurements. Therefore for 2010 we expect very competitive results
on the precision of the determined luminosity.

\section*{References}

\end{document}